\pdfoutput=1

\documentclass[11pt,twoside,a4paper,cmspaper,final,collab]{cms-tdr}
\def\svnVersion{227782}\def\cmsCernNo{2013-234}\def\cmsCernDate{\today}\def\cmsMessage{Published in the Journal of High Energy Physics as \href{http://dx.doi.org/10.1007/JHEP02(2014)024}{\doi{10.1007/JHEP02(2014)024.}}}

\begin{document}\cmsNoteHeader{TOP-12-007}

\hyphenation{had-ron-i-za-tion}
\hyphenation{cal-or-i-me-ter}
\hyphenation{de-vices}
\RCS$HeadURL: svn+ssh://svn.cern.ch/reps/tdr2/papers/TOP-12-007/trunk/TOP-12-007.tex $
\RCS$Id: TOP-12-007.tex 222062 2013-12-29 16:02:23Z alverson $
\newcommand{\usedLumi}{\ensuremath{5.3\fbinv}}
\providecommand{\eepm}{\ensuremath{\Pep\Pem}}
\providecommand{\mmpm}{\ensuremath{\Pgmp \Pgmm}}
\providecommand{\empm}{\ensuremath{\Pe^\pm \Pgm^\mp}}
\providecommand{\mll}{\ensuremath{M_{\ell\ell}}}

\cmsNoteHeader{TOP-12-007}
\title{Measurement of the \ttbar\ production cross section in the dilepton channel in pp collisions at $\sqrt{s} =8$\TeV}

\date{\today}

\abstract{The top-antitop quark ($\ttbar$) production cross section is measured in proton-proton collisions at
$\sqrt{s} = 8\TeV$ with the CMS experiment at the LHC, using a data sample  corresponding to an integrated luminosity of 5.3\fbinv. The measurement is performed by analysing  events with a pair of electrons or muons, or one electron and one muon, and at least two jets, one of which is identified as originating from hadronisation of a bottom quark. The  measured cross section is $239 \pm 2\stat \pm 11\syst\pm6\lum\unit{pb}$, for an assumed top-quark mass of 172.5\GeV, in agreement with the prediction of the standard model.}

\hypersetup{%
pdfauthor={CMS Collaboration},%
pdftitle={Measurement of the t t-bar production cross section  in the dilepton channel in pp collisions at
sqrt(s)=8 TeV},%
pdfsubject={CMS},%
pdfkeywords={CMS, physics, top physics}}

\maketitle

\section{Introduction}
\label{sec:introduction}

A precise measurement of the
\ttbar\
production cross section can be used to test the theory of quantum chromodynamics (QCD)
at next-to-next-to-leading-order (NNLO) level. It can be also used in global  fits
of the parton distribution
functions (PDF) at NNLO, and
allows an estimation  of  $\alpha_s(M_Z)$  as described in~\cite{CMSalphas7,czakonpdf}.
Furthermore, top-quark production is an important source of background in many searches for
physics beyond the standard model (SM). A large sample of top-quark events has been collected at the Large Hadron
Collider (LHC),
and studies of top-quark production have been conducted  in various decay channels
as well as searches for deviations from the SM
predictions~\cite{CMStopPublication2,top12001,ATLAStopPublication5, CMSpolar7,
CMStopPublication6,ATLAStopPublication2, ATLAStopPublication1}.

This paper presents
a measurement of the \ttbar\ production cross section, $\sigma_{\ttbar}$, based on the dilepton channel
(\eepm, \mmpm, and \empm)
 in a data sample of proton-proton collisions at $\sqrt{s} = 8\TeV$
corresponding to an integrated luminosity of \usedLumi\ recorded by the Compact Muon Solenoid (CMS) experiment.
In the SM, top quarks are predominantly produced in \ttbar\ pairs via the strong
interaction and decay almost exclusively to a \PW\ boson and a bottom quark.
We measure the \ttbar\ production
cross section selecting final states that contain two leptons of
opposite electric charge, momentum imbalance associated to the neutrinos
from the \PW\ boson decays, and two jets of particles resulting from the
hadronisation of two b quarks.

\section{The CMS detector and simulation}
\label{sec:detector}

The CMS detector~\cite{JINST} has a superconducting solenoid
occupying the central region
that provides
an axial magnetic field of 3.8\unit{T}. The silicon pixel
and the strip tracker cover $0 < \phi <2\pi$ in azimuth
and $\abs{\eta}<2.5$ in pseudorapidity, where $\eta$ is defined
as $\eta =-\ln[\tan(\theta/2)]$, with $\theta$ being the polar
angle
measured with respect to the
anticlockwise-beam direction. The lead-tungstate crystal electromagnetic
calorimeter and the brass/scintillator hadron calorimeter
are located inside the solenoid. Muons are measured in gas-ionisation
detectors embedded in the steel flux return yoke outside the
solenoid. The detector is nearly hermetic,
thereby providing reliable measurement of momentum imbalance
in the plane transverse to the beams. A two-tier trigger system
selects the most interesting \Pp\Pp\ collisions for offline analysis.

Several
MC event generators are used to simulate signal and background
events: \MADGRAPH~(v. 5.1.4.8)~\cite{madgraph}, \POWHEG~(r1380)~\cite{powheg} and \PYTHIA~(v. 6.424)~\cite{pythia},
depending on the process considered. The \MADGRAPH generator with spin correlations is used to model \ttbar\ events
with a top-quark mass of 172.5\GeV and combined with  \PYTHIA
to simulate   parton showering, hadronisation, and  the underlying event. The \MADGRAPH generator is also used
to simulate the \PW+jets and Drell--Yan (DY) processes. Single-top-quark  events are simulated using \POWHEG.
Inclusive production of the \PW\cPZ\ and \cPZ\cPZ\ diboson final states is simulated with \PYTHIA.
Production of \PW\PW\, fully leptonic final states is simulated with \MADGRAPH.
Decays of $\tau$ leptons are handled with \TAUOLA~(v. 2.75)~\cite{tauola}.
The contributions from \PW\PW, \PW\cPZ\, and \cPZ\cPZ\ (referred to as ``VV'')
and single-top-quark production are taken from MC simulations with
appropriate next-to-leading order (NLO) cross sections.
All other backgrounds are estimated from control samples extracted from collision data.

The \ttbar\ production cross section amounts to
$\sigma_{\ttbar} =  252.9^{+6.4}_{-8.6}\,\text{(scale)}\pm11.7\,\text{(PDF}+\alpha_s)\unit{pb}$
as calculated with the \textsc{Top++} program~\cite{top++} at
NNLO in perturbative QCD, including soft-gluon resummation at next-to-next-to-leading-log order~\cite{mitov},
and assuming a top-quark mass  $m_{\cPqt}= 172.5\GeV$.
The first uncertainty comes from the independent variation of the
factorisation and renormalisation scales, $\mu_F$ and $\mu_R$,
while the second one is associated to variations in the PDF and $\alpha_s$ following the
PDF4LHC prescriptions~\cite{pdf4lhcInterim}.
Expected signal yields in figures and tables are normalised to that value unless otherwise stated.

The simulated samples include additional interactions per bunch crossing (pileup),
with the distribution matching that observed in data.

\section{Event selection}
\label{sec:event_selection_yields}

Event selection is similar to that used
for the measurement of
the \ttbar\ dilepton cross section at $\sqrt{s} = 7\TeV$~\cite{top12001}.
At  trigger level, events are
required to have two electrons, two muons, or one electron and one muon,
where one of these leptons has  transverse momentum  $\pt > 17\GeV$ and the other has $\pt > 8\GeV$.
Events are then selected with two oppositely charged leptons
reconstructed
with the CMS particle-flow (PF) algorithm~\cite{PFPAS},
both with $\pt > 20\GeV$ and
$\abs{\eta} < 2.5$ for  electrons
and $\abs{\eta} <  2.1$ for  muons.
In events with more than one pair of leptons passing these selections, the pair
of  opposite-sign leptons with the largest value of total transverse momentum
is selected.
Events with $\tau$ leptons contribute to the measurement
only if they decay to electrons or muons that
satisfy the selection requirements.
The efficiency for
dilepton triggers is measured in data
through triggers based on transverse momentum imbalance.
The trigger efficiency is approximately 90\% to 93\% for the three final states.
Using the measured dilepton trigger efficiency in data,
the corresponding efficiencies in the simulation are corrected
by $\pt$ and $\eta$ multiplicative data-to-simulation scale factors (SFs),
which have an average value of $0.96$ and uncertainties in the range 1 to 2\%.

Charged-lepton candidates from W-boson decays are usually isolated from other particles
in the event. For each electron or muon candidate, a cone of $\Delta R <0.3$
is constructed around the track direction at the event vertex,
where $\Delta R$ is defined as $\Delta R = \sqrt{\smash[b]{(\Delta \eta)^2 + (\Delta \phi)^2}}$, and
$\Delta \eta$ and $\Delta \phi$ are the differences in pseudorapidity and azimuthal angle
between any  energy deposit and the axis of the lepton track.
  The scalar
sum of the $\pt$  of all particles reconstructed
with the PF algorithm, consistent with the
chosen primary vertex and contained within the cone, is calculated,
excluding the contribution from the lepton candidate itself.
The relative isolation discriminant, $I_\text{rel}$,
is defined as the ratio of this sum to the $\pt$
of the lepton candidate. The neutral component is
corrected for pileup based on the average energy density
deposited by  neutral particles in the event:
an average  transverse energy due
to pileup is determined event by event and is subtracted from the transverse energy in the
isolation cone. A lepton candidate is
rejected
if  $I_\text{rel}>0.15$.
The efficiency of the lepton selection is measured using a ``tag-and-probe'' method
in dilepton events enriched  in Z-boson candidates, as described in~\cite{inclusWZ3pb,top12001}.
The measured values for the combined identification and isolation
efficiencies are typically  of 96\% for muons and 90\% for electrons.
 Based on a comparison of lepton selection efficiencies in data and simulation,
the event yield in simulation is corrected
by $\pt$-and $\eta$-dependent SFs,
which have an average value
of 0.99 and uncertainties in the range 1 to 2\%
to provide consistency with data.
Considering also the  dilepton trigger,
the combined factors  have an average value
of 0.96 and uncertainties around 2\%
for the three  \ttbar\ final states.

Dilepton candidate events with an invariant mass $\mll < 20\GeV$ ($\ell=\Pe$ or $\mu$) are removed
to suppress backgrounds from heavy-flavour
resonances, as well as contributions from low-mass  DY processes.
Events with dilepton invariant masses within $\pm$15\GeV\ of the Z mass are also
rejected in the same-flavour channels.

Jets are reconstructed from the PF particle candidates
using the  anti-\kt clustering algorithm~\cite{antikt} with a distance parameter of
0.5.
The jet energy is corrected
for pileup in a manner similar to the correction of the energy inside the  lepton
isolation cone. Jet energy corrections are also applied as a function of the jet
$\pt$ and $\eta$~\cite{JESPUB}.
Events are required to have at least two reconstructed jets with $\pt> 30\GeV$ and $\abs{\eta}< 2.5$.

The missing transverse energy, $\ETslash$,
is defined as the magnitude of the momentum imbalance, which is the negative
 sum of the momenta of all reconstructed particles in the plane transverse to the beams.
A value of $\ETslash>40\GeV$ is required in the \eepm\ and \mmpm\ channels while no \ETslash\ requirement is imposed
for the \empm\ mode,
as there is very little contamination from DY events in this channel.

Since
\ttbar\ events contain jets from  hadronisation of \cPqb\ quarks,
requiring their presence can
reduce background from events without b quarks.
Jets are identified as  b jets using the combined secondary vertex algorithm
(CSV)~\cite{BTV-11-004-pub}. The operating point chosen for CSV corresponds to an
identification efficiency
of about 85\% and a  misidentification (mistag) probability of about 10\%~\cite{BTV-12-DP} for
light-flavour jets (u, d, s and gluons).
The selection requires the presence  of at least one b jet in the event.

\begin{figure}[htbp]
\centering
\includegraphics[width=0.48\textwidth]{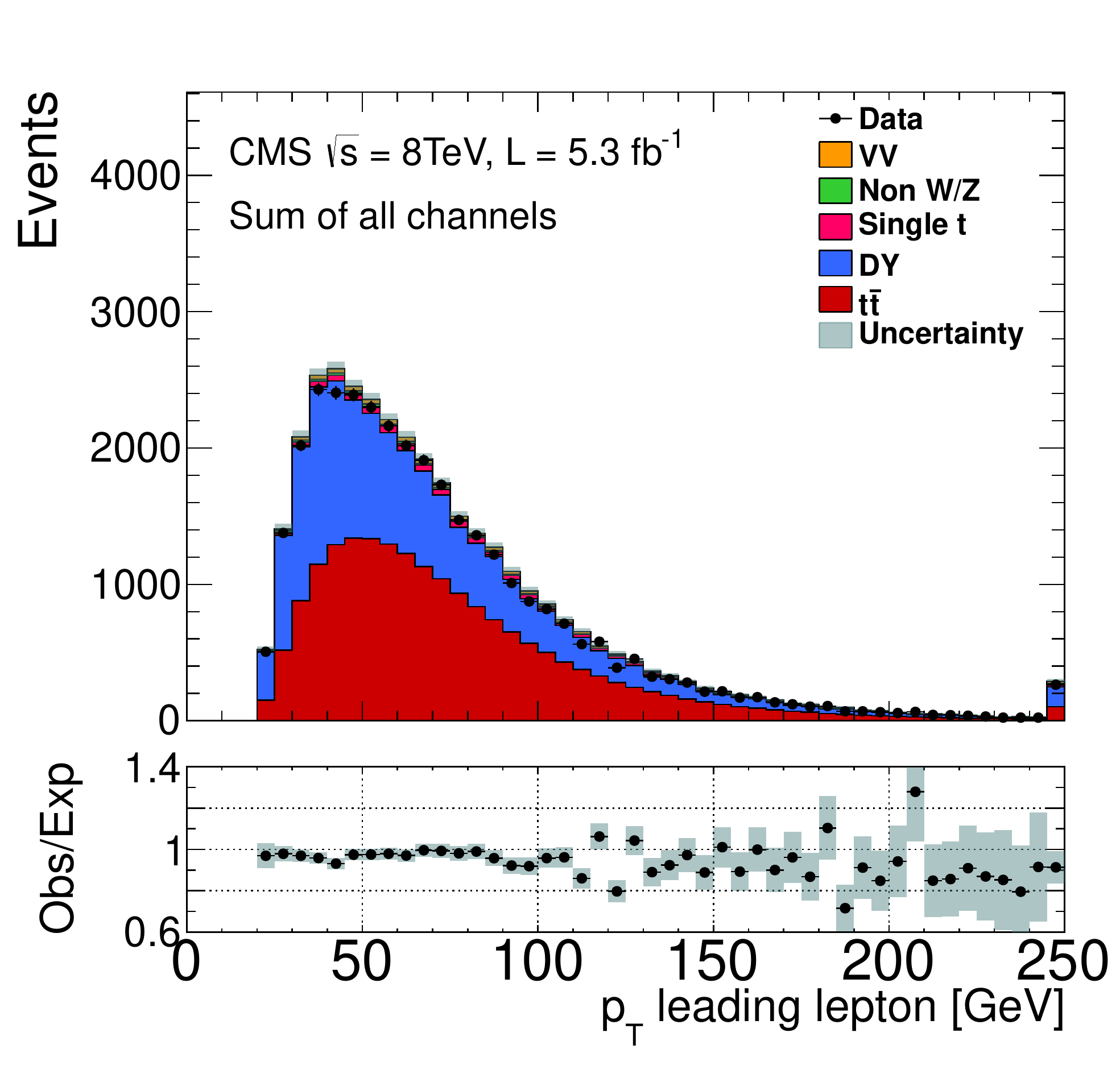}
\includegraphics[width=0.48\textwidth]{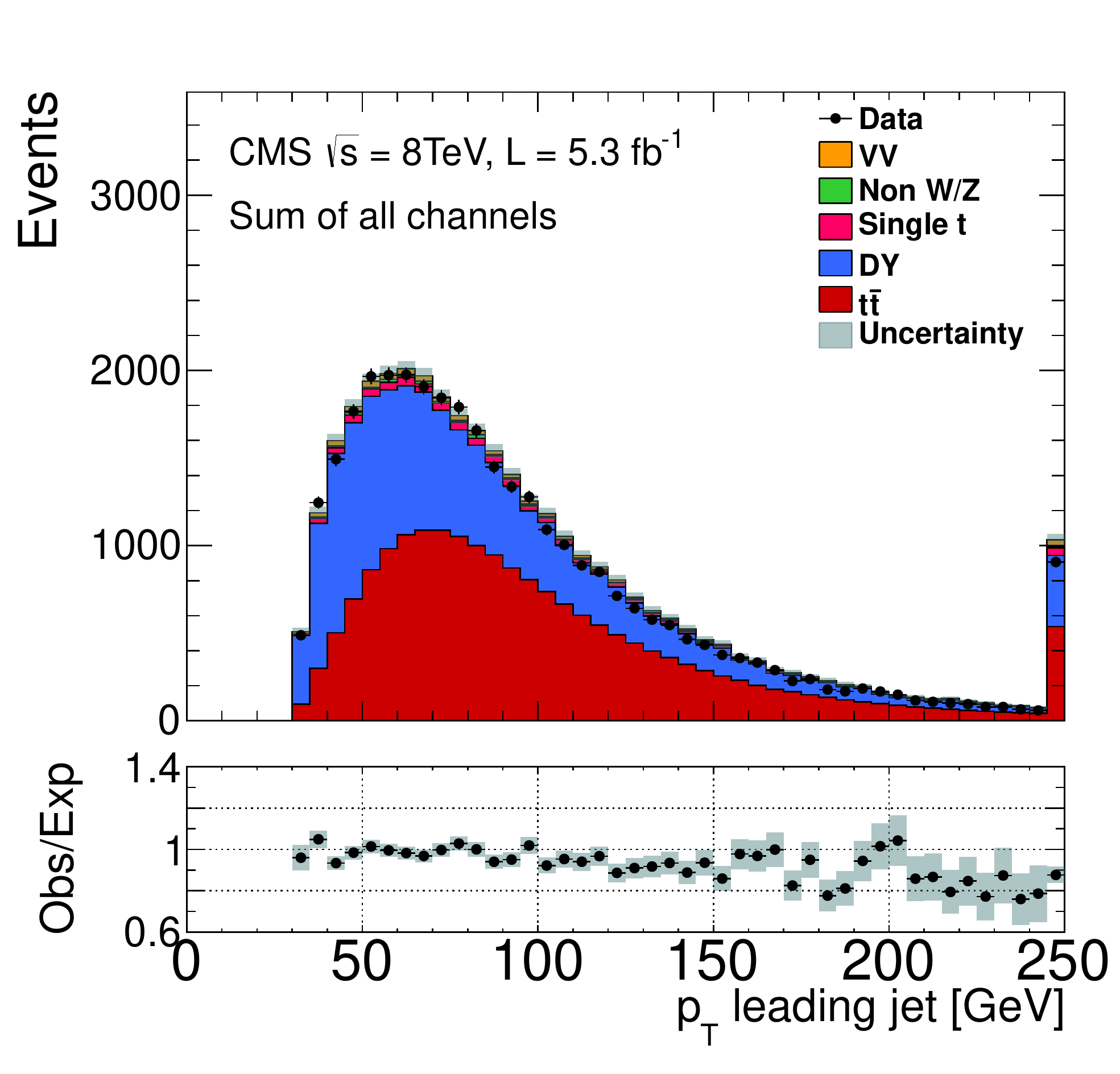}
\caption{The $\pt$ distributions of the highest-$\pt$
lepton (left) and
jet (right) after the jet multiplicity selection, for all three final states.
 The expected distributions for \ttbar\ signal and individual backgrounds are
 shown after data-based corrections are applied; the last bin contains the overflow events.
The hatched bands correspond to the total
statistical uncertainty in the  event yields for the sum of the
\ttbar\ and background predictions. The ratios of data to the sum of the
 expected yields are given at the bottom.
}
\label{fig:pt1}
\end{figure}

\begin{figure}[htbp]
\centering
\includegraphics[width=0.48\textwidth]{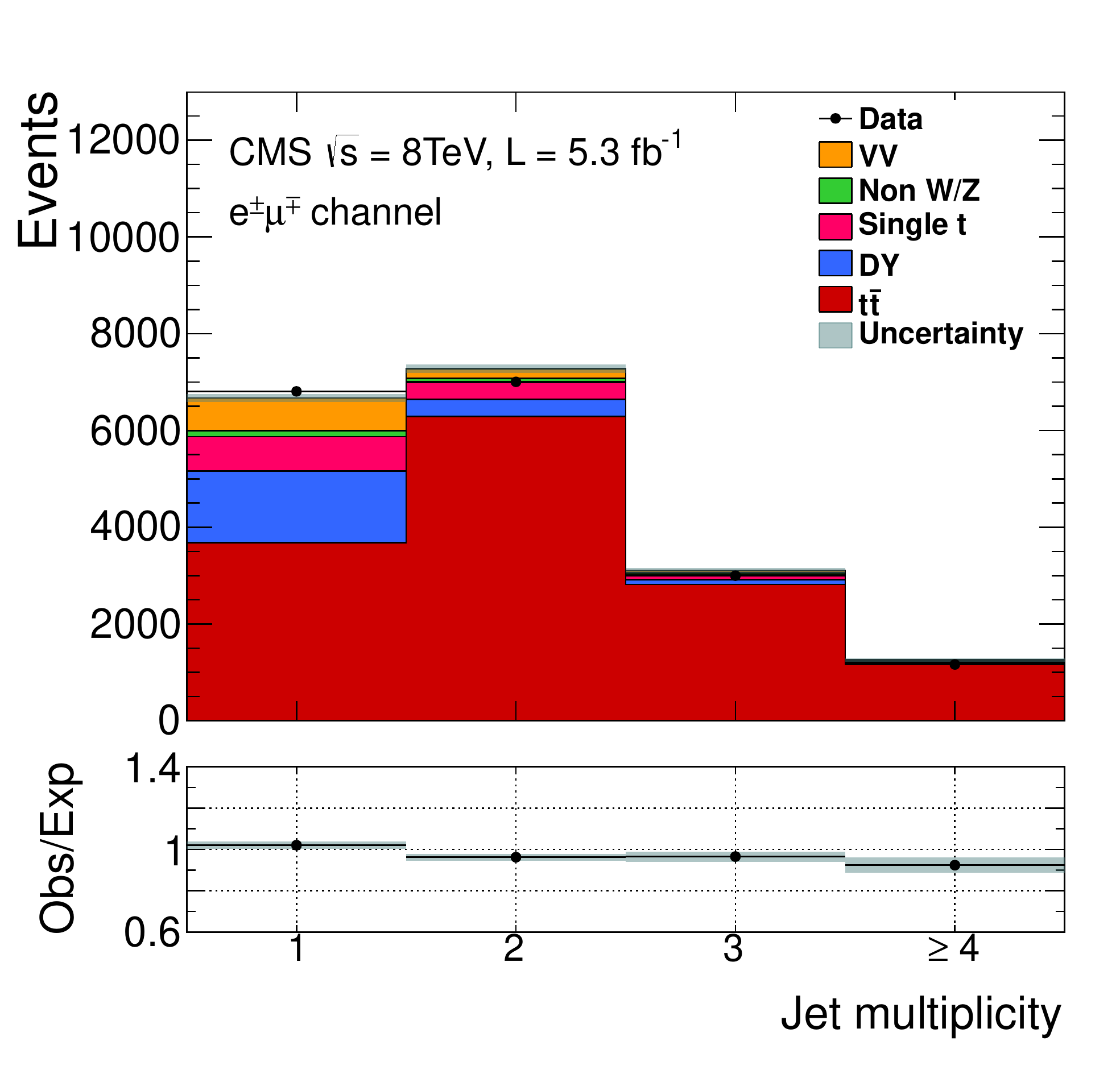}
\includegraphics[width=0.48\textwidth]{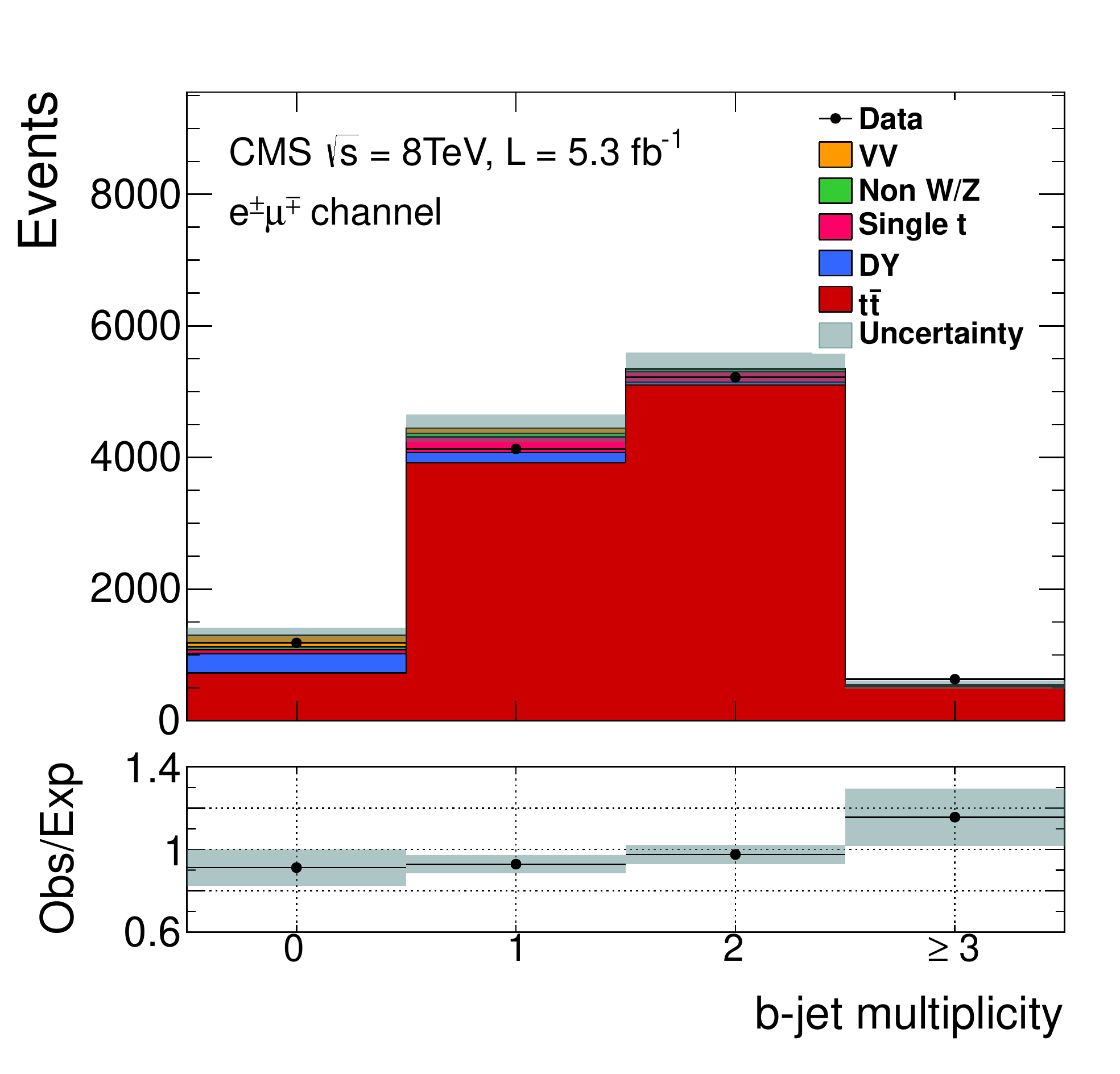}
\caption{Jet multiplicity (left) in events passing the dilepton criteria,
and (right) b-jet multiplicity in events passing the full event
selections but before the b-jet requirement, for the \empm\ channel.
In the right figure, the hatched bands show the total statistical and
b-jet systematic uncertainties in the  event yields for the sum of the \ttbar\ and background predictions.
The hatched bands in the left figure show only the total statistical uncertainty on the predicted event yields.
The ratios of data to the sum of the expected yields are given at the bottom.
}
\label{fig:njetsDD_nbtags}
\end{figure}

Figure~\ref{fig:pt1} shows the $\pt$ distributions of the highest-$\pt$
lepton and jet
after jet multiplicity selection, for all three final states combined.
In this and the following figures the signal yields refer to an assumed top-quark mass of
172.5\GeV.
The hatched regions correspond to the total statistical uncertainties
in the predicted event yields.
The ratio of the data to the sum of simulations and data-based predictions for the signal and backgrounds
is shown in the bottom panels.
A detailed description of the different background estimates
is given in section~\ref{sec:bkgd}.
The multiplicities of selected jets
and
 b jets
are
shown in figure~\ref{fig:njetsDD_nbtags} for the \empm\ channel,
 which is expected to have less background contamination.
A similar level of agreement is obtained with the \eepm\ and \mmpm\ channels.

\section{Background determination}
\label{sec:bkgd}
Backgrounds in this analysis arise from single-top-quark, DY
and VV events,
in which at least two
prompt leptons are produced from Z or W decays. Other background sources,
such as
\ttbar\
or W+jets events with decays into lepton+jets
and where at least one jet is incorrectly reconstructed
as a lepton (which mainly happens for electrons) or a lepton from the
decay of bottom or charm
hadrons (which mainly happens for muons), are grouped into the non-W/Z lepton category.
Background yields from single-top-quark  and VV events are estimated from simulation,
while all other backgrounds
are estimated from
data.

The DY background is estimated using the ``$R_\text{out/in}$'' method~\cite{top12001,top10001,CMStopPublication2}
in which the  events  outside of the Z mass
window are obtained by normalising the event yield from simulation to the observed number of events
inside the Z mass window.
The data-to-simulation scale factor
is found to be
 $1.3\pm0.4$ for the \empm\ channel. This value is compatible with $1.5\pm0.5$, which is estimated
using a template fit as described
in~\cite{top12001}. For the \eepm\ and \mmpm\ channels the
factors
are found to be $1.7\pm0.5$ and $1.6\pm0.5$, respectively.

Non-prompt leptons can arise from decays of mesons or heavy-flavour quarks,
jet misidentification, photon conversions, or finite resolution detector
effects whereas prompt leptons usually originate from decays of W or Z bosons and
are isolated and well identified.
Backgrounds with non-prompt leptons are estimated~\cite{CMSHWWlegacy} from a control
sample of collision data in which leptons are selected with relaxed  identification
and isolation requirements  defining the loose lepton candidate, while
the set of signal selection cuts described in section~\ref{sec:event_selection_yields}
 defines the tight
lepton candidate. The prompt and non-prompt lepton ratios are defined as
the ratio of the number of tight candidates to the number of loose ones as
measured from samples enriched in leptonic decays of Z bosons or in
QCD dijet events, respectively. These ratios, parametrized as a function
of $\pt$ and $\eta$ of the lepton, are then used to weight the events
in the loose-loose dilepton sample, to obtain the estimated contribution
from the non-prompt lepton background in the signal region.
The systematic uncertainty comes from the jet $\pt$ spectrum
in dijet events and amounts, together with the statistical one, to 40\% of the estimated yield.

\section{Sources of systematic uncertainty}
\label{sec:systematics}

Simulated events are scaled according to the lepton
efficiency correction factors,
which are typically close to one,
measured using control samples in data,
leading to a 1 to 2\% uncertainty
in the  \ttbar\  selection efficiency.

The impact of uncertainty in the jet energy scale (JES) and jet energy resolution (JER) are estimated from
the change observed
in the number of selected MC \ttbar\ events after varying
the jet momenta
within the JES uncertainties~\cite{JESPUB}, and in the case
of JER
by an $\eta$-dependent correction with an average of $\pm$10\%.
For the \eepm\ and \mmpm\ channels these uncertainties are also propagated to $\ETslash$
resulting in a larger uncertainty  than for the \empm\ channel.

The uncertainties
on the b jet scale factors in \ttbar\ signal events are approximately 2\% for b jets and
10\% for
mistagged jets~\cite{BTV-11-004-pub,BTV-12-DP}, depending on the $\pt$ of the jets. They are propagated
to the \ttbar\ selection
efficiency in simulated events.

The uncertainty assigned to the pileup
simulation amounts to 0.8\%, as obtained by varying the inelastic cross section by 5\%.
The uncertainty in the integrated luminosity is 2.6\%~\cite{lumiPAS13}.

The systematic effects related to the missing higher-order diagrams
in  \MADGRAPH  are estimated with two different methods.
The uncertainty in the signal acceptance
is determined by varying the
renormalisation
 and factorisation  scales
simultaneously up and down by a factor of two
using \MADGRAPH, and the uncertainty is taken as the maximum difference after the final event selection.
The effect
on the calculated
\ttbar\ production cross section is 2.3\%, which is the value used in the analysis for this uncertainty.
This estimate is cross-checked by comparing the predictions of the leading-order and NLO
generators \MADGRAPH and  \POWHEG, where both use \PYTHIA for hadronisation and extra
radiation. The systematic uncertainty is found to be 2.2\%,
comparable with the above estimate.

The matching between the matrix elements (ME) and the parton shower (PS) evolution
is done by applying the MLM prescription~\cite{MEPS}.
Changing the thresholds that control the matching of partons from
the matrix element with those from PS by factors of 0.5
and 2.0 for one of the parameters
(minimum \kt  measure between partons) and 0.75 and 1.5 for
the other (jet matching threshold for the \kt-MLM scheme)
compared to the default thresholds,
produces a 1.6\% variation in the
\ttbar\ event selection efficiency.

The uncertainty arising from the hadronisation model
affects mainly the JES and the  fragmentation
of b jets. As the b-jet efficiencies
and mistagging rates are taken from data, no additional uncertainty
is expected from this source.
The uncertainty in the JES already
contains a contribution from the uncertainty in the hadronisation.
The hadronisation uncertainty is also determined by comparing samples
 of events generated with \POWHEG  where the hadronisation is modelled
 with   \PYTHIA
or \HERWIG, and the effect on the calculated
\ttbar\ cross section is 1.4\%, which  is well within the JES uncertainty.

Uncertainties in the selected number of single-top-quark and VV events
are calculated following the same prescription as for the signal yield.
In addition, an uncertainty in the cross sections for single-top-quark and
VV backgrounds, taken from measurements and estimated to
be approximately 20\%~\cite{CMSWWZZPublication8, CMSWWPublication7, CMSWWHiggsPublication7,
CMSWWZZPublication7, CMSZZPublication7, CMStopPublicationstop4, ATLASWWPublication,
ATLASWZPublication,ATLASZZPublication}, is added in quadrature.

Table~\ref{tab:breakdown_comb}  summarizes the magnitude of the
systematic uncertainties on the \ttbar\ production cross section from the different
sources.

\begin{table}[htb]
\centering
\topcaption{Summary of the individual contributions to the systematic uncertainty on
the $\sigma_{\ttbar}$  measurement.
The uncertainties are given in pb.
The statistical uncertainty on the result is given for comparison.}

\begin{tabular}{lccc}
\hline
Source                        &  \eepm  & \mmpm & \empm  \\ \hline
\hline
Trigger efficiencies          &  4.1  &  3.0    &  3.6 \\
Lepton efficiencies           &  5.8  &  5.6    &  4.0 \\
Lepton energy scale           &  0.6  &  0.3    &  0.2 \\
Jet energy scale              &  10.3 &  10.8   &  5.2 \\
Jet energy resolution         &  3.2  &  4.0    &  3.0 \\
b-jet tagging                     &  1.9  &  1.9    &  1.7 \\
Pileup                        &  1.7  &  1.5    &  2.0 \\
Scale  ($\mu_F$ and $\mu_R$)          &  5.7  &  5.5    &  5.6 \\
Matching partons to showers   &  3.9  &  3.8    &  3.8 \\
Single top quark              &  2.6  &  2.4    &  2.3 \\
VV                            &  0.7  &  0.7    &  0.5 \\
Drell--Yan                     &  10.8 &  10.3   &  1.5 \\
Non-W/Z leptons               &  0.9  &  3.2    &  1.9 \\
\hline
Total systematic             &  18.6 &  18.6   &  11.4 \\
\hline
Integrated luminosity         &  6.4  &  6.1    &  6.2 \\
\hline
Statistical                   &  5.2  &  4.5    &  2.6 \\
\hline
\end{tabular}
\label{tab:breakdown_comb}
\end{table}

\section{Results}
\label{sec:measure}

The \ttbar\ production cross section is
measured by counting events after applying the selection criteria
described in  section~\ref{sec:event_selection_yields}.
Table~\ref{tab:yieldsChannels} shows the total number of events observed in
data and the number of signal and background events expected from simulation
or estimates from data.
Table~\ref{tab:xsecChannels} lists the mean acceptance
(which contains contributions from $\PW\rightarrow$ $\tau \nu_{\tau}$, with leptonic $\tau$
decays)
multiplied by the selection efficiency
and  the branching fraction
in the dilepton final state,
and the measured cross section for each of the three final states, \eepm, \mmpm,
and \empm, which give compatible results.
The \eepm\ and \mmpm\  channels have two additional sources of uncertainty,
arising from the DY background estimation and  from the propagation of the JES to the
$\ETslash$ estimation, which limit the
precision of the measurement of  $\sigma_{\ttbar}$ in those final states.

A combination of the three final states
using the BLUE method~\cite{BLUE}  yields a measured cross section of
$\sigma_{\ttbar} = 239.0\pm2.1\stat\pm 11.3\syst\pm6.2\lum\unit{pb}$
for a top-quark mass of 172.5\GeV.
In the combination, the systematic uncertainties are 100\% correlated across channels,
except those associated to the lepton efficiencies,
which have a correlation coefficient
of 0.64 for $\eepm$ with $\empm$ and 0.55 for $\mmpm$ with $\empm$.
Finally, the uncertainties associated with the data-based estimates
and the statistical uncertainties are taken as uncorrelated.

In this analysis the dependence of the acceptance on the top-quark mass is found to be quadratic
within the present uncertainty of
the top-quark mass~\cite{PDG2012}.
The cross-section dependence in the range 160--185\GeV can be parametrized as
\begin{equation}
\sigma_{\ttbar}/\sigma_{\ttbar}\,(m_{\cPqt}  = 172.5) = 1.00 - 0.009\times (m_{\cPqt}  -172.5) - 0.000168 \times (m_{\cPqt}  -172.5)^2
\end{equation}
where $m_{\cPqt}$ is given in \GeVns.
Assuming a top-quark mass value of 173.2\GeV~\cite{PDG2012}, a cross section value
$\sigma_{\ttbar} = 237.5\pm13.1\unit{pb}$
is obtained.

\begin{figure}[htbp]
\centering
\includegraphics[width=0.48\textwidth]{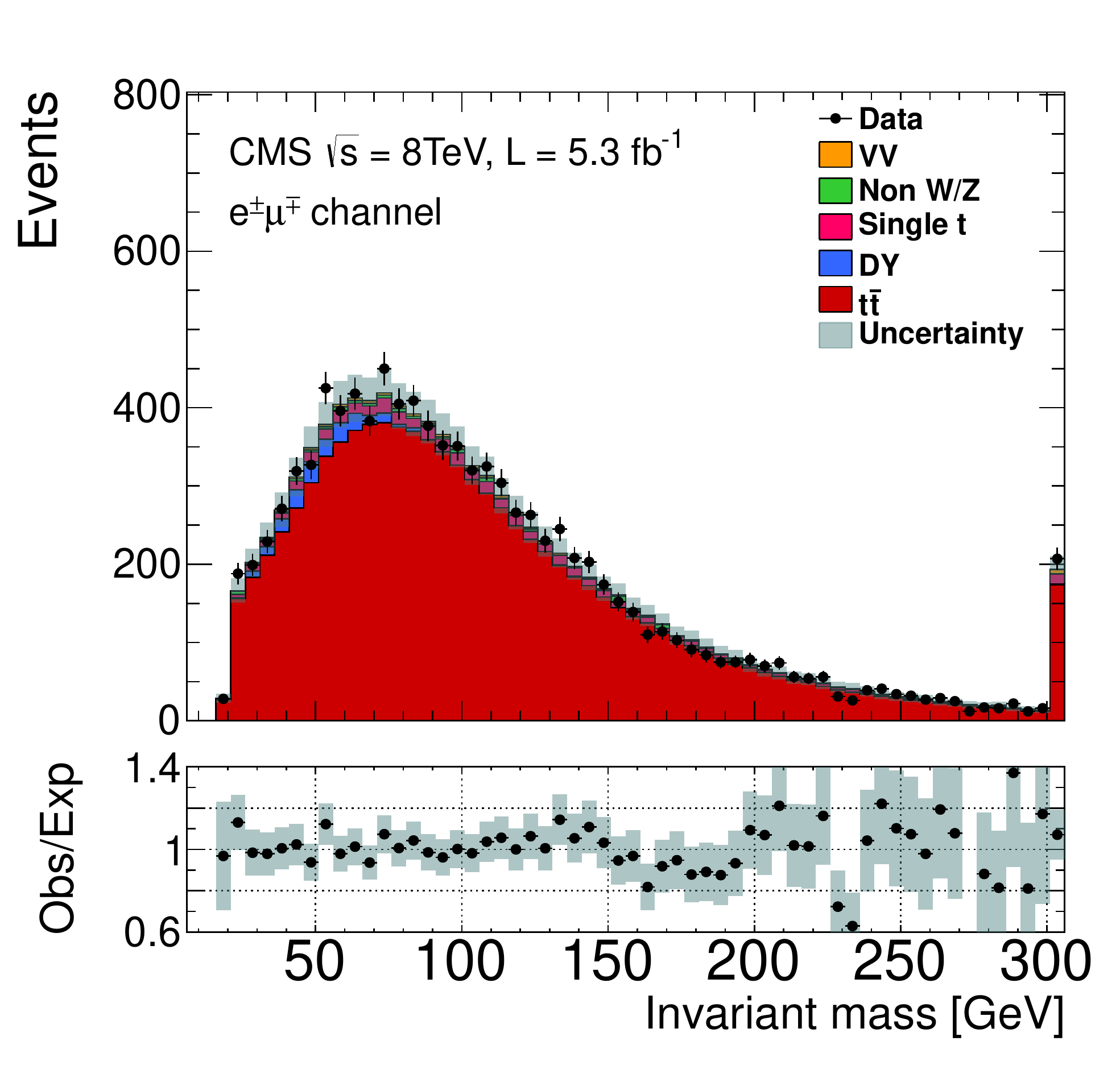}
\includegraphics[width=0.48\textwidth]{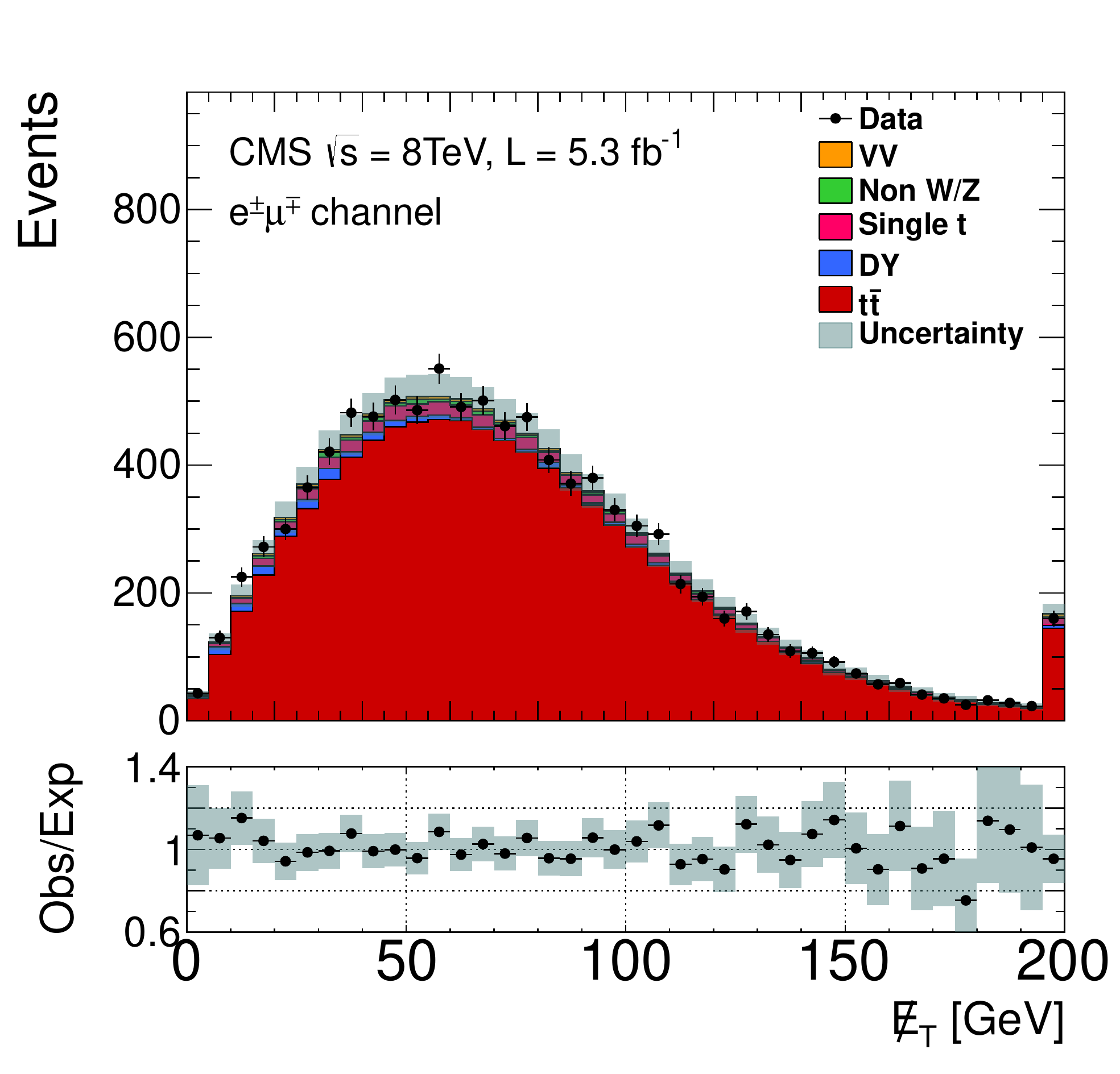}
\includegraphics[width=0.48\textwidth]{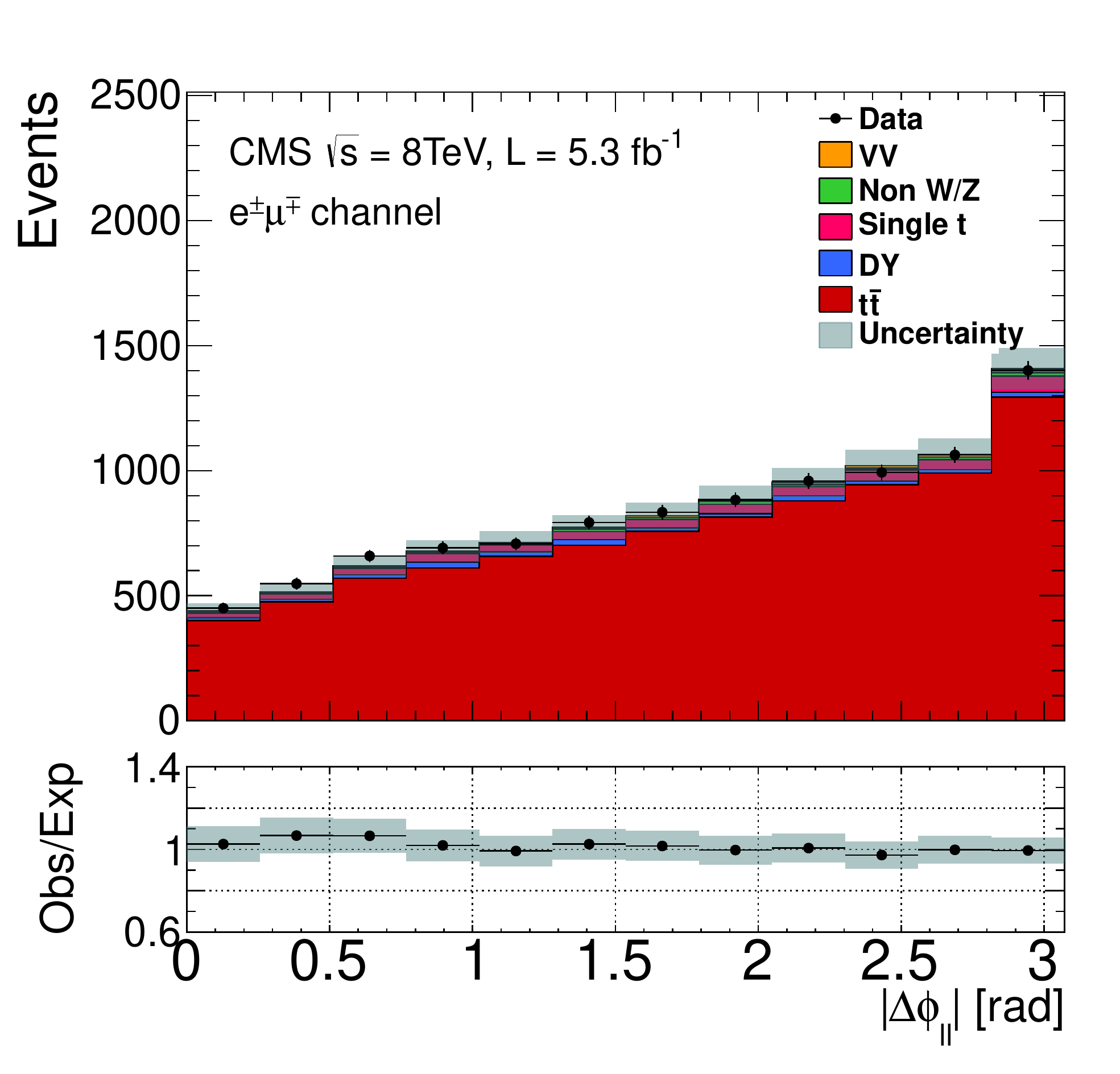}
\caption{Distributions of (upper left) the dilepton invariant-mass, (upper right) the \ETslash,  and (lower)
the difference of the azimuthal angle between the two selected leptons,
after
the b-jet multiplicity selection and for the \empm\ channel.
For the first two plots the last bin contains the overflow events.
 The expected distributions for \ttbar\ signal, in this case, are normalised to the measured
 \ttbar\ cross section.
The hatched bands correspond to the total
 uncertainty in the predicted event yields for the sum of the
\ttbar\ and background predictions.
The ratios of data to the sum of the expected yields are given at the bottom.
}
\label{fig:invmass_met}
\end{figure}

Figure~\ref{fig:invmass_met} shows the distributions of
\mll, \ETslash\ and the difference of the azimuthal angle between the two selected leptons
($\Delta\phi_{\ell\ell}$)  and their ratios to expectations
for the \empm\ channel, which dominates  the combination.

\section{Summary}
\label{sec:conclusions}
A measurement of the \ttbar production cross section in proton-proton collisions
at $\sqrt{s} =8\TeV$ is presented for events containing
a lepton pair (\eepm, \mmpm,  \empm),
at least two jets
with at least one tagged as b jet,
and a large imbalance in  transverse momentum in the final state.
The measurement is obtained through an event-counting
analysis based on a data sample corresponding to
\usedLumi.
The result obtained by combining the three final states
is $\sigma_{\ttbar} = 239\pm2\stat\pm
11\syst\pm6\lum\unit{pb}$,
in agreement with the prediction of the
 standard model for a top-quark mass of 172.5\GeV.

\begin{table}[!htb]
\centering
\topcaption{Number of dilepton events after applying the event selection and requiring at
least one b jet.
The results are given for the individual sources of background, \ttbar\ signal
with a top-quark mass of 172.5\GeV  and $\sigma_{\ttbar}
= 252.9\unit{pb}$, and data. The uncertainties correspond to the statistical and systematic components added in quadrature.
}
\begin{tabular}{lr@{}c@{}lr@{}c@{}lr@{}c@{}l}
\hline
  &  \multicolumn{9}{c}{Number of events} \\Source   &
  \multicolumn{3}{c}{\eepm} & \multicolumn{3}{c}{\mmpm} & \multicolumn{3}{c}{\empm}      \\\hline
Drell--Yan                 & 386  & $\pm$ & 116 & 492  & $\pm$ & 148 & 194  & $\pm$ & 58  \\
Non-W/Z leptons           & 25   & $\pm$ & 10  & 114  & $\pm$ & 46  & 185  & $\pm$ & 72  \\ \hline
Single top quark          & 127  & $\pm$ & 28  & 157  & $\pm$ & 34  & 413  & $\pm$ & 88  \\
VV                  & 30   & $\pm$ & 8   & 39   & $\pm$ & 10  & 94   & $\pm$ & 21  \\ \hline
Total background          & 569  & $\pm$ & 120 & 802  & $\pm$ & 159 & 886  & $\pm$ & 130 \\ \hline
\ttbar\ dilepton signal   & 2728 & $\pm$ & 182 & 3630 & $\pm$ & 250 & 9624 & $\pm$ & 504 \\ \hline
Data  &  \multicolumn{3}{c}{3204}  & \multicolumn{3}{c}{4180}  & \multicolumn{3}{c}{9982}  \\ \hline
\end{tabular}
\label{tab:yieldsChannels}
\end{table}

\begin{table}[ht]
\centering
\topcaption{
The total efficiencies $\epsilon_\textrm{total}$, i.e. the products of event acceptance, selection efficiency and branching fraction for the respective \ttbar\ final states, as estimated from simulation for a top-quark mass of 172.5\GeV, and the measured \ttbar\ production cross sections, where the uncertainties are from statistical, systematic and integrated luminosity components, respectively.
}
\begin{tabular}{lccc} \hline
 &  \eepm &  \mmpm & \empm \\ \hline
$\epsilon_\text{total}$ (\%)  & 0.203 $\pm$ 0.012
                     & 0.270 $\pm$ 0.017
                     & 0.717 $\pm$ 0.033 \\ \hline

$\sigma_{\ttbar}$ (pb) & 244.3 $\pm$ 5.2 $\pm$ 18.6 $\pm$ 6.4
                       & 235.3 $\pm$ 4.5 $\pm$ 18.6 $\pm$ 6.1
                       & 239.0 $\pm$ 2.6 $\pm$ 11.4 $\pm$ 6.2 \\ \hline
\end{tabular}
\label{tab:xsecChannels}
\end{table}

\section*{Acknowledgments}

We congratulate our colleagues in the CERN accelerator departments for the excellent performance of the LHC and thank the technical and administrative staffs at CERN and at other CMS institutes for their contributions to the success of the CMS effort. In addition, we gratefully acknowledge the computing centres and personnel of the Worldwide LHC Computing Grid for delivering so effectively the computing infrastructure essential to our analyses. Finally, we acknowledge the enduring support for the construction and operation of the LHC and the CMS detector provided by the following funding agencies: BMWF and FWF (Austria); FNRS and FWO (Belgium); CNPq, CAPES, FAPERJ, and FAPESP (Brazil); MES (Bulgaria); CERN; CAS, MoST, and NSFC (China); COLCIENCIAS (Colombia); MSES and CSF (Croatia); RPF (Cyprus); MoER, SF0690030s09 and ERDF (Estonia); Academy of Finland, MEC, and HIP (Finland); CEA and CNRS/IN2P3 (France); BMBF, DFG, and HGF (Germany); GSRT (Greece); OTKA and NIH (Hungary); DAE and DST (India); IPM (Iran); SFI (Ireland); INFN (Italy); NRF and WCU (Republic of Korea); LAS (Lithuania); CINVESTAV, CONACYT, SEP, and UASLP-FAI (Mexico); MBIE (New Zealand); PAEC (Pakistan); MSHE and NSC (Poland); FCT (Portugal); JINR (Dubna); MON, RosAtom, RAS and RFBR (Russia); MESTD (Serbia); SEIDI and CPAN (Spain); Swiss Funding Agencies (Switzerland); NSC (Taipei); ThEPCenter, IPST, STAR and NSTDA (Thailand); TUBITAK and TAEK (Turkey); NASU (Ukraine); STFC (United Kingdom); DOE and NSF (USA).

Individuals have received support from the Marie-Curie programme and the European Research Council and EPLANET (European Union); the Leventis Foundation; the A. P. Sloan Foundation; the Alexander von Humboldt Foundation; the Belgian Federal Science Policy Office; the Fonds pour la Formation \`a la Recherche dans l'Industrie et dans l'Agriculture (FRIA-Belgium); the Agentschap voor Innovatie door Wetenschap en Technologie (IWT-Belgium); the Ministry of Education, Youth and Sports (MEYS) of Czech Republic; the Council of Science and Industrial Research, India; the Compagnia di San Paolo (Torino); the HOMING PLUS programme of Foundation for Polish Science, cofinanced by EU, Regional Development Fund; and the Thalis and Aristeia programmes cofinanced by EU-ESF and the Greek NSRF.

\bibliography{auto_generated}

\cleardoublepage \appendix\section{The CMS Collaboration \label{app:collab}}\begin{sloppypar}\hyphenpenalty=5000\widowpenalty=500\clubpenalty=5000\textbf{Yerevan Physics Institute,  Yerevan,  Armenia}\\*[0pt]
S.~Chatrchyan, V.~Khachatryan, A.M.~Sirunyan, A.~Tumasyan
\vskip\cmsinstskip
\textbf{Institut f\"{u}r Hochenergiephysik der OeAW,  Wien,  Austria}\\*[0pt]
W.~Adam, T.~Bergauer, M.~Dragicevic, J.~Er\"{o}, C.~Fabjan\cmsAuthorMark{1}, M.~Friedl, R.~Fr\"{u}hwirth\cmsAuthorMark{1}, V.M.~Ghete, C.~Hartl, N.~H\"{o}rmann, J.~Hrubec, M.~Jeitler\cmsAuthorMark{1}, W.~Kiesenhofer, V.~Kn\"{u}nz, M.~Krammer\cmsAuthorMark{1}, I.~Kr\"{a}tschmer, D.~Liko, I.~Mikulec, D.~Rabady\cmsAuthorMark{2}, B.~Rahbaran, H.~Rohringer, R.~Sch\"{o}fbeck, J.~Strauss, A.~Taurok, W.~Treberer-Treberspurg, W.~Waltenberger, C.-E.~Wulz\cmsAuthorMark{1}
% ...
\vskip\cmsinstskip
% ...
1:~~Also at Vienna University of Technology, Vienna, Austria\\
2:~~Also at CERN, European Organization for Nuclear Research, Geneva, Switzerland\\
% ...

\end{sloppypar}
\end{document}